\documentclass[11pt,a4paper,oneside,groupcitations]{article}
\usepackage[T1]{fontenc}
\usepackage[ansinew]{inputenc}
\usepackage[english]{babel}
\usepackage{amsfonts}
\usepackage{amsmath}
\usepackage{bm}
\usepackage{array}
\usepackage{amsthm}
\usepackage{amssymb}
\usepackage{graphicx}
\usepackage{braket}
\usepackage{verbatim}
\usepackage[table]{xcolor}
\usepackage{caption}
\usepackage{cite}
\usepackage{textcomp}
\usepackage{url}
\usepackage{multicol}
\newcommand{\be}{\begin{equation}}
\newcommand{\ee}{\end{equation}}
\definecolor{pinegreen}{rgb}{0.0, 0.47, 0.44}

\theoremstyle{definition}

\theoremstyle{remark}

\title{\bf On variable-order fractional linear viscoelasticity} %
\author{
Andrea~Giusti$^{a,b}$
,
Ivano~Colombaro$^{c}$,
Roberto~Garra$^{d}$,\\
Roberto~Garrappa$^{e}$,
$\ $and
Andrea~Mentrelli$^{b,f,g}$
\\
\\
$^a$ {\em Institute for Theoretical Physics, ETH Zurich}
\\
{\em Wolfgang-Pauli-Strasse 27, 8093 Zurich, Switzerland}
\\
\\
$^b${\em Alma Mater Research Center on Applied Mathematics (AM$^2$)}
		\\
{\em Via Saragozza 8, 40123 Bologna, Italy}
\\
\\
$^c${\em Faculty of Engineering, Free University of Bozen-Bolzano}
\\
{\em Piazza Universit\`a 5, 39100 Bolzano, Italy}
\\
\\
$^d$ {\em Section of Mathematics, International Telematic University Uninettuno}
\\
{\em Corso Vittorio Emanuele II 39, 00186 Rome, Italy}
\\
\\
$^e${\em Department of Mathematics, University of Bari}
\\
{\em Via E. Orabona 4, 70125 Bari, Italy}
\\
\\
$^f${\em Department of Mathematics, University of Bologna}
	\\
	{\em Piazza di Porta San Donato 5, 40126 Bologna, Italy} 
\\
\\
$^g${\em I.N.F.N., Sezione di Bologna, I.S. FLAG}
	\\
	{\em viale B. Pichat 6/2, 40127 Bologna, Italy}
}
\begin{document}
\maketitle
\begin{abstract}
We discuss a generalization of fractional linear viscoelasticity based on Scarpi's approach to variable-order fractional calculus. After reviewing the general mathematical framework, we introduce the {\em variable-order fractional Maxwell model} as a simple example for our analysis. We then provide some physical considerations for the fractionalisation procedure and on the choice of the transition functions. Lastly, we compute the material functions for the considered model and evaluate them numerically for exponential-type and Mittag-Leffler-type order functions.
\end{abstract}
\newpage
\section{Introduction}
\label{sec:1}
\setcounter{equation}{0}
Physical phenomena displaying long memory ({\em i.e.}, non-exponential relaxation) and non-local effects have been attracting increasing attention in both mathematical and physical communities for the past three decades. Within this research field, fractional calculus turns out to be probably the most fashionable approach when it comes to the mathematical modelling of a wide class of effects featuring non-localities in space and/or time. 

Notably, fractional calculus finds one of its major physical motivations in rheology. As discussed in \cite{MSB}, in order to provide a theoretical justification for Nutting's law \cite{Nutting} G. W. Scott Blair and collaborators proposed (a version of) the constitutive relation (see \cite{SB}, and references therein)
$$
\sigma (t) = \kappa_\alpha \, D_t ^\alpha \epsilon (t) \, , \quad \alpha \in (0,1) \, ,
$$
with $D_t ^\alpha$ denoting a ``precursor'' of the Caputo fractional derivative. The main feature of this model, which is known in the literature as the {\em Scott Blair model} of viscoelasticity, is that it is somewhat intermediate between the behaviour of an Hookean elastic body ($\alpha = 0$) and a viscous Newtonian substance ($\alpha = 1$). For a complete account of the role of Fractional Calculus in viscoelasticity we refer the interest reader to \cite{MainardiSpada} and the monograph \cite{MainardiBook}. 

Fractional calculus is a branch of mathematics that aims at extending the notions of derivative and repeated integration to arbitrary real, or even complex, orders (see {\em e.g.}, \cite{Samko1993fractint,MainardiGorenflo1997FC}).
The standard procedure leading to the standard definitions of fractional operators relies on the analytic continuation of the Cauchy formula for repeated integration that leads to the so-called Riemann-Liouville fractional integral \cite{MainardiGorenflo1997FC}. A fractional derivative is then defined as the left-inverse operator to the Riemann-Liouville integral (this condition known as the {\em fundamental theorem of fractional calculus} \cite{diethelm2020why}). If we take this requirement as the defining property of fractional derivatives then there exist several distinct operators that satisfy this condition. For instance, both the Riemann-Liouville and the Caputo derivative satisfy the fundamental theorem of fractional calculus with respect to the Riemann-Liouville integral \cite{MainardiGorenflo1997FC}. If we then consider generalisations of Riemann-Liouville fractional integral as starting point for the formulation of a generalised fractional calculus then the main features of fractional operators are \cite{diethelm2020why,Hanyga2020}: (i) the fact that such operators are convolution-type integro-differential operators with {\em weakly-singular kernels}; (ii) generalised fractional integrals and derivatives must satisfy the fundamental theorem of fractional calculus. These properties can then be recast as conditions on the integral kernels for these operators, see {\em e.g.}, \cite{Luchko-1,Luchko-2,Luchko-3}. 

A natural question that comes to mind then is ``what happens if the order of a fractional operator is a function of space and/or time?''. This opens up the broader subject of {\em variable-order fractional calculus} (for a partial overview on this topic we refer the reader to \cite{Samko2013,Ortigueiraetal2019,Sunetal2019}). In this work we shall focus on a recently revived proposal for the definition of variable-order fractional derivative known as {\em Scarpi derivative} (originally presented in \cite{Scarpi1972a,Scarpi1972b}). In particular, we will employ the main results of \cite{GarrappaGiustiMainardi2021, GarrappaGiusti2023}, where Scarpi's approach has been framed within the framework of the theory of general fractional operators (see {\em e.g.},\cite{Luchko-1,Luchko-2,Luchko-3}). Furthermore, we will take advantage of the computational methods for 
exponential-type variable-order fractional differential equations involving the Scarpi derivative developed in \cite{GarrappaGiusti2023}.

This work is therefore organised as follows. In Sec.~\ref{sec:2} we discuss the generalities of Scarpi's approach to variable-order fractional calculus. In Sec.~\ref{sec:3} we investigate the properties of the Maxwell model of linear viscoelasticity, as a toy example. In Sec.~\ref{sec:massacro} we present a critical assessment of the shortcomings of the previous literature \cite{cinesata} on the application of Scarpi's variable-order fractional calculus in viscoelasticity. Lastly, in Sec.~\ref{sec:conc} we present some concluding remarks.
\section{Variable-order fractional operators: Scarpi's approach}
\label{sec:2}
\setcounter{equation}{0}
In this section we briefly summarise the basics of Scarpi's approach to variable-order fractional calculus. This part of the work is based on \cite{GarrappaGiustiMainardi2021,GarrappaGiusti2023}, hence we refer the interested readers to these pivotal studies for further details.

Let $\alpha \, : \, [0, +\infty) \, \longrightarrow \, (0,1)$ be a (bounded) function such that:
\begin{itemize}
\item[{\em (i)}] $\alpha$ is locally integrable on $[0, +\infty)$.
\item[{\em (ii)}] The Laplace transform of $\alpha (t)$, {\em i.e.},
$$
A(s) = \mathcal{L} \left( \alpha (t) \, ; \, s \right) = 
\int_0 ^\infty {\rm e}^{- s t} \, \alpha (t) \, {\rm d} t \, ,
$$
exists and its analytical expression is known explicitly.
\item[{\em (iii)}] The following condition holds:
$$
\lim _{t \to 0^+} \alpha (t) = \bar{\alpha} \in (0, 1) \, .
$$
\end{itemize}
Then, we define the fractional integral operator
\be
I^{\alpha (t)} f(t) := \int _0 ^t \psi _\alpha (t - \tau) \, f(\tau) \, {\rm d} \tau \, , \,\,\, \mbox{with} \,\,\, \psi _\alpha (t) := \mathcal{L}^{-1} \left( s^{-s A(s)} \, ; \, t \right) \, ,
\ee
and the associated Caputo-type fractional derivative as
\be
D^{\alpha (t)} f(t) := \int _0 ^t \phi _\alpha (t - \tau) \, f'(\tau) \, {\rm d} \tau \, , \,\,\, \mbox{with} \,\,\, \phi _\alpha (t) := \mathcal{L}^{-1} \left( s^{s A(s)-1} \, ; \, t \right) \, ,
\ee 
where $f$ is a sufficiently regular function on $[0, +\infty)$ (see \cite{GarrappaGiustiMainardi2021,GarrappaGiusti2023}), $f'$ denotes the first derivative of $f$ with respect to its variable and $\mathcal{L}^{-1}$ indicates the inverse Laplace transform operation.

As proven in \cite{GarrappaGiustiMainardi2021}, the kernels $\psi _\alpha (t)$ and $\phi _\alpha (t)$ form a Sonine pair, which implies that
\be
D^{\alpha (t)} I^{\alpha (t)} f(t) = f(t) \quad \mbox{and} \quad
I^{\alpha (t)} D^{\alpha (t)} f(t) = f(t) - f(0^+) \, ,
\ee
for sufficiently regular functions (see \cite{GarrappaGiustiMainardi2021,GarrappaGiusti2023}). Lastly, it is easy to show that 
\be
\mathcal{L}\Big( D ^{\alpha(t)} f(t) \, ; \, s \Big) = s^{sA(s)}F(s) - s^{sA(s)-1} f(0^+) \, ,
\ee
and 
\be
\mathcal{L}\Big( I ^{\alpha(t)} f(t) \, ; \, s \Big) = s^{-sA(s)}F(s) \, ,
\ee
where $F(s):= \mathcal{L} \left(f(t) \, ; \, s \right)$. We refer again the reader to \cite{GarrappaGiustiMainardi2021} for detailed proofs of these important results.
\subsection{Exponential-type transition functions}
\label{sec:2-1}
In this work we will focus on a specific class of variable-order exponents $\alpha (t)$ known as {\em exponential-type transition functions}. This type of evolution laws for the order of a fractional derivative have been originally introduced and discussed in detail in \cite{GarrappaGiusti2023}. We therefore refer the interested reader to \cite{GarrappaGiusti2023} for details on both theoretical and numerical aspects of variable-order operators involving these transition functions. In the following we simply summarise some key results that will be then employed in the following sections of this work.

Let $0<\alpha_1<1 $ and $0<\alpha_2<1$, and consider an order function $\alpha (t)$ on $[0,+\infty)$ such that
\be \label{eq:alpha-exp-trans}
\alpha(t) = \alpha_2 + (\alpha_1 - \alpha_2) \text{e}^{-ct} \, ,
\ee
where $c>0$ is known as {\em transition rate}. Such a transition function starts-off at $\alpha (0^+) = \alpha_1$ and smoothly approaches $\alpha (+\infty) = \alpha _2$ at late times.

The Laplace transform of $\alpha (t)$ in \eqref{eq:alpha-exp-trans} is easily computed with standard methods and yields
\be
A(s) = \frac{\alpha_2 c + \alpha_1 s}{s(c+s)} \,, \quad \Re (s) >0 \, .
\ee
Then, the Laplace transform of the kernel $\Psi_{\alpha_1, \alpha_2} (s)$ of the associated Scarpi fractional integral reads
\be
\Psi_{\alpha_1, \alpha_2} (s) = s^{-sA(s)} = s^{- \frac{\alpha_2 c + \alpha_1 s}{c+s}} \, ,
\ee
while the kernel for the fractional derivative, in the Laplace domain, is given by
\be
\Phi_{\alpha_1, \alpha_2} (s) = s^{sA(s)-1} = s^{-\frac{(1-\alpha_2)c + (1-\alpha_1)s}{c+s}} \,.
\ee
\subsection{Mittag-Leffler-type transition functions}
\label{sec:2-2}
Another type of transition function that we shall consider in the following rely on the Mittag-Leffler function \cite{MLfunction, MLmaina} (see {\em e.g.}, \cite{MLgen} for its role in the generalisation of fractional calculus). Specifically, let $0<\alpha_1<1 $, $0<\alpha_2<1$, $\beta>0$, $c>0$, and 
\be
\label{eq:alpha-ML-trans}
\alpha (t) = \alpha_2 + (\alpha_1 - \alpha_2) E_\beta (-c \, t^\beta) \, , \qquad t>0 \, ,
\ee
where 
$$
E_\beta (t) := \sum _{k=0} ^\infty \frac{t^k}{\Gamma (\beta k + 1)}
$$
denotes the (one-parameter) Mittag-Leffler function. 

Eq.~\eqref{eq:alpha-ML-trans} describes a function that starts-off at $\alpha_1$ and continuously approaches $\alpha_2$ at late time. Its Laplace transform is easily computed and reads
\be
A(s) = \frac{\alpha_2 c + \alpha_1 s^\beta}{s(c+s^\beta)} \,, \quad |s|>|c|^{1/\beta} \, .
\ee
The corresponding integral kernels for Scarpi's operators, in the Laplace domain, then read
\be
\Psi_{\alpha_1, \alpha_2} (s) = s^{-sA(s)} = s^{- \frac{\alpha_2 c + \alpha_1 s^\beta}{c+s^\beta}} \quad \mbox{and} \quad 
\Phi_{\alpha_1, \alpha_2} (s) = s^{sA(s)-1} = s^{-\frac{(1-\alpha_2)c + (1-\alpha_1)s^\beta}{c+s^\beta}} \, .
\ee
\section{Variable-order Fractional Maxwell model}
\label{sec:3}
\setcounter{equation}{0}
Over the past decade, a significant effort has been devoted to the study of applications of variable-order fractional operators in applied sciences. In particular, as an example we refer to \cite{liu} and references therein.

In this work we wish to provide an implementation of a {\em variable-order fractional linear theory of viscoelasticity} based on Scarpi's approach to fractional calculus, which has recently been given new life in \cite{GarrappaGiustiMainardi2021}. To this end, we will discuss a generalisation of the {\em Maxwell model} as a toy example. The generalisation of other models of linear viscoelasticity, preserving the linearity of the model, can the be carried out following a similar procedure to the one discussed here.

The standard (ordinary) {\em Maxwell model} is defined in terms of the constitutive law (see {\em e.g.}, \cite{MainardiSpada, MainardiBook})
\be
\sigma +\frac{\eta}{E}\, \sigma ' = \eta \, \epsilon ' \, , 
\ee
where $\sigma = \sigma (t)$ and $\epsilon = \epsilon (t)$ are the {\em stress} and {\em strain} functions, respectively, $\eta$ denotes the viscosity coefficient, $E$ is the elastic modulus, and the prime denotes the derivative with respect to time.

This model is easily generalised to the well-known {\em fractional Maxwell model} (see {\em e.g.}, \cite{MainardiSpada, MainardiBook}), defined in terms of the constitutive equation
\be
\sigma + a_1 D^{\alpha} \sigma =  b_1 D^{\alpha} \epsilon \, , \qquad \alpha \in (0,1) \, , \quad a_1, b_1 > 0 \, ,
\ee
where $D ^{\alpha}$ denotes either the (constant-order) Riemann-Liouville fractional derivative or the (con\-stant-order) Caputo fractional derivative. Employing either of the two operators does not lead to significantly different physical behaviours for the system, and it mostly relates to the analysis of the initial conditions for both the stress and strain \cite{MainardiSpada, MainardiBook}. The key difference is however in the dimensions of the (constant) coefficients $a_1$ and $b_1$. Indeed, since $[D^{\alpha}] = (\mbox{time})^{-\alpha}$, then $[a_1] = (\mbox{time})^{\alpha}$ and $[ b_1 \, D^{\alpha} \epsilon] = [\sigma]$. This implies $[b_1] = [\sigma] (\mbox{time})^{\alpha}$, since the strain is dimensionless ({\em i.e.}, $[\epsilon] = 1$). To avoid this explicit dependence on the fractional power in the coefficients of the constitutive law, we observe that $a_1, b_1 \propto T^\alpha$, with $T$ a {\em characteristic time-scale} for the model. Hence, we can always write $a_1 = \bar{a}_1 \, T^\alpha$ and $b_1 = \bar{b}_1 \, T^\alpha$, with $\bar{a}_1$ being a dimensionless positive real constant and $\bar{b}_1$ denoting a coefficient with the dimensions of an elastic modulus. If we then reabsorb $T$ in the definition of $D^\alpha$, {\em via} a time redefinition $t \to \bar{t} =t/T$, then the {\em reparametrised fractional Maxwell model} reads
\be
\label{eq:reptwo}
\sigma + \bar{a}_1 \bar{D}^{\alpha} \sigma =  \bar{b}_1 \bar{D}^{\alpha} \epsilon \, , \qquad \alpha \in (0,1) \, , \quad \bar{a}_1, \bar{b}_1 > 0 \, ,
\ee
with $\bar{D}^{\alpha}$ denoting a fractional derivative with respect to the dimensionless time $\bar{t}$, $[\bar{a}_1] =1$, and $[\bar{b}_1] = [\sigma]$.
{\bf From this point on we will adopt this second representation ({\em i.e.}, the one provided in Eq.~\eqref{eq:reptwo}) for our operators and constitutive relations and we will drop the ``overlines'' to lighten the notation.} 

The above reformulation of the model in terms of a dimensionless time allows us to perform an unambiguous variable-order generalisation of the model \eqref{eq:reptwo} that {\bf preserves its linearity}. Specifically, starting from
Eq.~\eqref{eq:reptwo} we define the constitutive relation of the {\bf variable-order fractional Maxwell model} as
\be
\label{eq:scarpi-maxwell}
\sigma + {a}_1 \, {D}^{\alpha (t)} \sigma = {b}_1 \, {D}^{\alpha (t)} \epsilon \, ,  \quad {a}_1, {b}_1 > 0 \, ,
\ee
with $t$ denoting the dimensionless time, $[{a}_1] = 1$, $[{b}_1]=[\sigma]$, and $\alpha (t)$ being a function on $[0, + \infty)$ such that $0<\alpha (t)<1$ for all $t \in [0, + \infty)$ and satisfying the conditions {\em (i--iii)} in Sec.~\ref{sec:2}.

Following standard procedures (see {\em e.g.}, \cite{MainardiSpada, MainardiBook, GSvisco, Pipkin}) we can now compute the {\em creep compliance} $J(t)$ and {\em relaxation modulus} $G(t)$ for our system. In detail, assuming $a_1 \sigma (0^+) = b_1 \epsilon (0^+)$, since we are considering a Caputo-type fractional derivative (for further details on this matter we refer the reader to \cite[Appendix B]{MainardiSpada} and to \cite{MainardiBook,GSvisco}), we have that the Laplace domain representation of Eq.~\eqref{eq:scarpi-maxwell} reads
\be
\widetilde{\sigma} (s) = s \, \widetilde{G} (s) \, \widetilde{\epsilon} (s) \, ,
\ee
with
\be
\label{eq:G-LT}
\widetilde{G} (s) = b_1 \, \frac{s^{s A(s) - 1}}{1 + a_1 \, s^{s A(s)}} \, ,
\ee
where the tilde denotes the Laplace transform of the corresponding time-dependent functions. The Laplace transform of the creep compliance is then obtained taking advantage of a fundamental relation of linear viscoelasticity \cite{MainardiBook,GSvisco}, {\em i.e.},
$$
s^2 \, \widetilde{G} (s) \, \widetilde{J} (s) = 1 \, ,
$$
which yields
\be
\label{eq:G-LT-J}
\widetilde{J} (s) = \frac{1}{b_1} \, \frac{1 + a_1 \, s^{s A(s)}}{s^{s A(s) + 1}} \, .
\ee
The representation in time of the creep compliance and of the relaxation modulus is obtained by inverting the Laplace transforms $\widetilde{J} (s)$ and $\widetilde{G} (s)$, respectively. This procedure, for the case of variable-order operators, is however only possible {\em via} numerical methods, as detailed in \cite{GarrappaGiustiMainardi2021, GarrappaGiusti2023}.

Let us stress that $G(t)$ and $J(t)$, derived as above, generate the solutions of the {\em stress relaxation} (constant strain) and {\em creep} (constant stress) {\em experiments}. Indeed, let us consider, for instance, a constant-strain experiment with $\epsilon (t) = \epsilon_0$, with $\epsilon_0 \neq 0$ being a constant, $\forall t\geq0$. Plugging this condition into Eq.~\eqref{eq:scarpi-maxwell} yields
\be
\label{eq:constant-strain}
\sigma + {a}_1 \, {D}^{\alpha (t)} \sigma = 0 \, .
\ee
Taking the Laplace transform of Eq.~\eqref{eq:constant-strain} gives 
\be
\widetilde{\sigma} (s) + a_1 
\left[ s^{s A(s)} \, \widetilde{\sigma} (s) 
- s^{s A(s)-1} \, \sigma (0^+)\right] = 0 \, ,
\ee
that can be rewritten as
\be
\widetilde{\sigma} (s) = a_1 \, \sigma (0^+) \, 
\frac{s^{s A(s) - 1}}{1 + a_1 \, s^{s A(s)}}
= \frac{a_1 \, \sigma (0^+)}{b_1} \, \widetilde{G} (s) \, , 
\ee
namely
\be
\sigma (t) = \frac{a_1 \, \sigma (0^+)}{b_1} \, G(t) \, .
\ee
Similarly, one can show that for a creep experiment, {\em i.e.}, $\sigma (t) = \sigma_0 \neq 0$, $\forall t\geq0$, one finds
\be
\epsilon (t) = \sigma_0 \, J_\star (t) \, ,
\ee
where 
$$
\widetilde{J}_\star (s) = 
\frac{1}{b_1} \, \frac{1 + \frac{b_1 \epsilon(0^+)}{\sigma_0} \, s^{s A(s)}}{s^{s A(s) + 1}} \, .
$$
Then, if one assumes the constraint on the initial conditions $a_1 \sigma (0^+) = b_1 \epsilon (0^+)$, mentioned at the beginning of this section, then the response to a constant strain is given by $\sigma (t) = \epsilon_0 \, G(t)$, while the creep experiment yields $\epsilon (t) = \sigma_0 \, J(t)$.
\subsection{Physical motivation and discussion}
\label{sec:3-1}
The driving motivation for the study of variable-order fractional operators, beside a purely mathematical interest, lays in the fact that these operators have the capability of describing physical systems whose underlying structure evolves with time. This allows mathematical models based on such variable-order operators to adapt to the changing properties of the system. Although one could argue that this would represent just an emerging description of the system at large, a model that ``averages'' over the complexities of a system providing a reliable macroscopic description of the latter would still serve many practical purposes.

Coming to the choice of the transition functions $\alpha(t)$ in Eq.~\eqref{eq:alpha-exp-trans} and~\eqref{eq:alpha-ML-trans}, this was not random. Indeed, assuming that variable-order fractional calculus can be employed successfully to describe some physical systems, then this raises questions concerning the physical nature of the order function $\alpha (t)$. Physical intuition would suggest that $\alpha (t)$ should not be regarded as a mere best-fit tool that can be tuned at one's pleasure, but rather as a quantity associated to its own evolution equation. Such evolution equation would then deserve a fundamental justification to the same extent as the equation involving the variable-order operators acting on physical observables.

Eq.~\eqref{eq:alpha-exp-trans} is the solution of the {\em relaxation-type problem}, {\em i.e.}, 
\begin{equation*}
    \left\{
    \begin{aligned}
    & \alpha' (t) = - c \, \alpha (t) + c \, \alpha_2 \, , \\ 
    & \alpha (0^+) = \alpha_1 \, ,
    \end{aligned} \right.
\end{equation*}
whereas Eq.~\eqref{eq:alpha-ML-trans} is the solution of the {\em Caputo-type fractional relaxation-type problem} that reads
\begin{equation*}
    \left\{
    \begin{aligned}
    & {{}^{\rm C} \! D}^\beta \alpha (t) = - c \, \alpha (t) + c \, \alpha_2 \, , \\ 
    & \alpha (0^+) = \alpha_1 \, ,
    \end{aligned} \right.
\end{equation*}
where ${{}^{\rm C} \! D}^\beta$ denotes the Caputo fractional derivative of order $\beta\in(0,1)$ with respect to $t$. These evolution equations play a key role in several, radically different, physical scenarios. So, due to their ubiquity in physics these evolution equation have directed our focus toward transition functions of the form \eqref{eq:alpha-exp-trans} and \eqref{eq:alpha-ML-trans} for this work.
\subsection{Numerical evaluation of the material functions}
\label{sec:3-2}
As discussed in this section, the material functions $G(t)$ and $J(t)$ determine the stress relaxation and creep response of the system. Therefore it is of interest to study the qualitative behaviour of the variable-order Maxwell model \eqref{eq:scarpi-maxwell} given a legitimate expression for $\alpha (t)$ and some values for the model parameters by studying the numerical evaluation of the relaxation modulus and creep compliance.

In the following we will focus on the {\em dimensionless version} of the material functions, {\em i.e.},
\be
G_{\rm d} (t) := G(t)/ b_1 \quad \mbox{and} \quad
J_{\rm d} (t) := b_1 \, J (t) \, ,
\ee
thus providing a numerical evaluation of the {\em inverse Laplace transform} of
\be
\label{eq:Gd-LT}
\widetilde{G}_{\rm d} (s) = \frac{s^{s A(s) - 1}}{1 + a_1 \, s^{s A(s)}} 
\ee
and
\be
\label{eq:Jd-LT}
\widetilde{J}_{\rm d} (s) = \frac{1 + a_1 \, s^{s A(s)}}{s^{s A(s) + 1}} \, ,
\ee
by taking advantage of a {\em modified Talbot method} (see  \cite[Appendix A]{GarrappaGiustiMainardi2021} and \cite{{WeidemanTrefethen2007}}).

Furthermore, we shall also compare the behaviour of $G_{\rm d} (t)$ and $J_{\rm d} (t)$ for the variable-order fractional Maxwell model with the constant-order expressions
\be 
G^{(\alpha)} _{\rm M, d} (t) = \frac{1}{a_1} \, E_\alpha \left( - \frac{t^\alpha}{a_1} \right) 
\ee
and
\be 
J^{(\alpha)} _{\rm M, d} (t) = \frac{t^\alpha}{\Gamma (\alpha +1)} + a_1 \, , 
\ee
for $t>0$, $a_1>0$, and $\alpha\in (0,1)$, at early and late times. For the sake of precision, in the following plots we will restore the dimension of time $t$ via the replacement $t \to t/T$, where again $T$ is the characteristic timescale of the model as discussed above.

Additionally, here we further investigate a particular sub-case of the fractional Maxwell model, namely the Scott Blair model \cite{MainardiSpada,MainardiBook} obtained by setting $a_1 = 0$ in the constitutive law~\eqref{eq:scarpi-maxwell}. For this model we also compare the asymptotic behaviour, at early and late times, of the dimensionless material functions with the known expression for the constant-order Scott Blair model, that are easily computed and read
\be 
G^{(\alpha)} _{\rm SB, d} (t) = \frac{t^{-\alpha}}{\Gamma(1-\alpha)}
\ee
and
\be 
J^{(\alpha)} _{\rm SB, d} (t) = \frac{t^{\alpha}}{\Gamma(\alpha + 1)} \, ,
\ee
for $t>0$ and $\alpha\in (0,1)$.
\begin{figure}[ht!]
    \centering
    \includegraphics[scale=0.8]{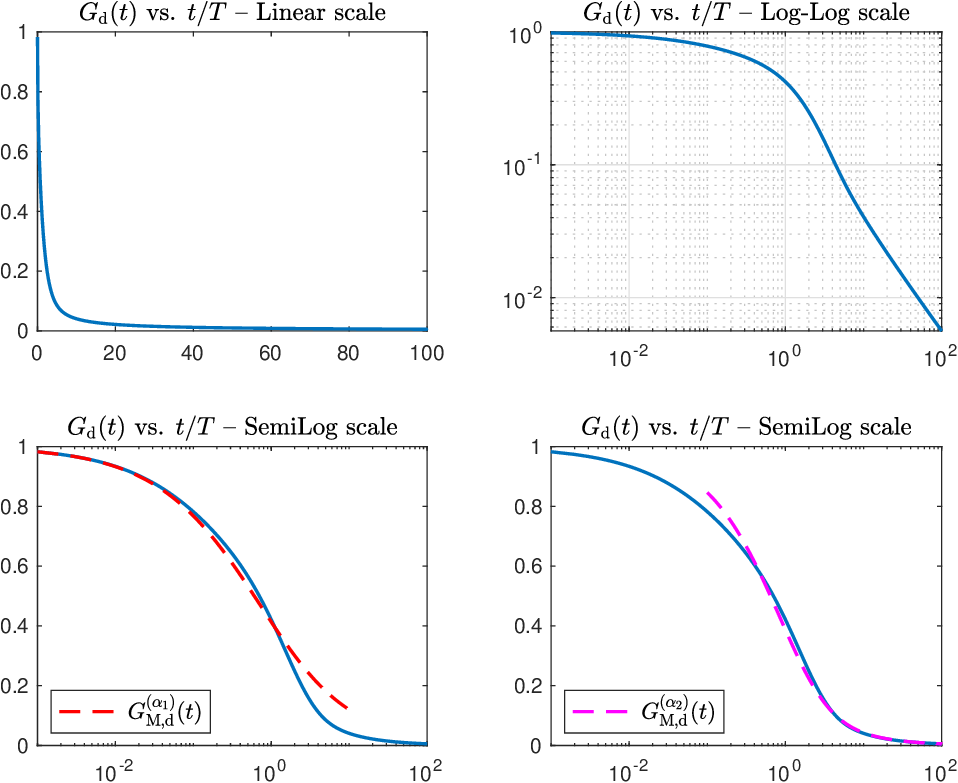}
    \caption{Dimensionless relaxation modulus ${G}_{\rm d}(t)$ [continuous line] as a function of $t/T$ for the variable-order fractional Maxwell model given an exponential-type transition $\alpha(t)$ (see Eq.~\eqref{eq:alpha-exp-trans}) with $\alpha_1 = 0.6$, $\alpha_2=0.8$, $c=2$, and $a_1 = 1$.}
    \label{fig:Gd-Maxwell}
\end{figure}
\begin{figure}[ht!]
    \centering
    \includegraphics[scale=0.8]{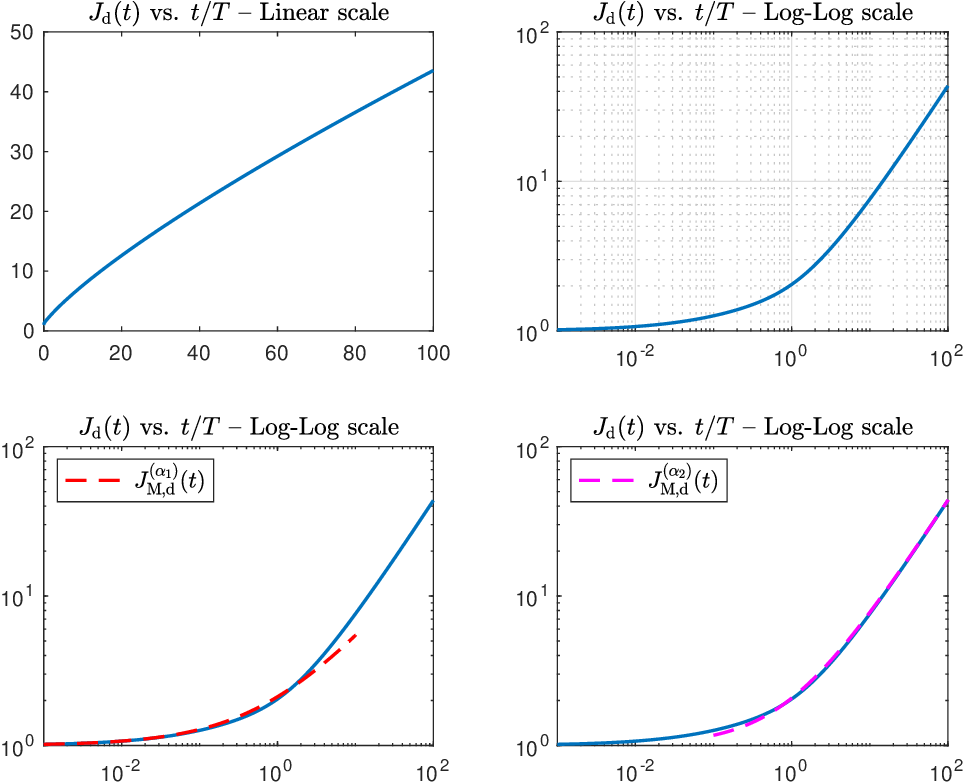}
    \caption{Dimensionless creep compliance ${J}_{\rm d}(t)$ [continuous line] as a function of $t/T$ for the variable-order fractional Maxwell model given an exponential-type transition $\alpha(t)$ (see Eq.~\eqref{eq:alpha-exp-trans}) with $\alpha_1 = 0.6$, $\alpha_2=0.8$, $c=2$, and $a_1 = 1$.}
    \label{fig:Jd-Maxwell}
\end{figure}
\begin{figure}[ht!]
    \centering
    \includegraphics[scale=0.4]{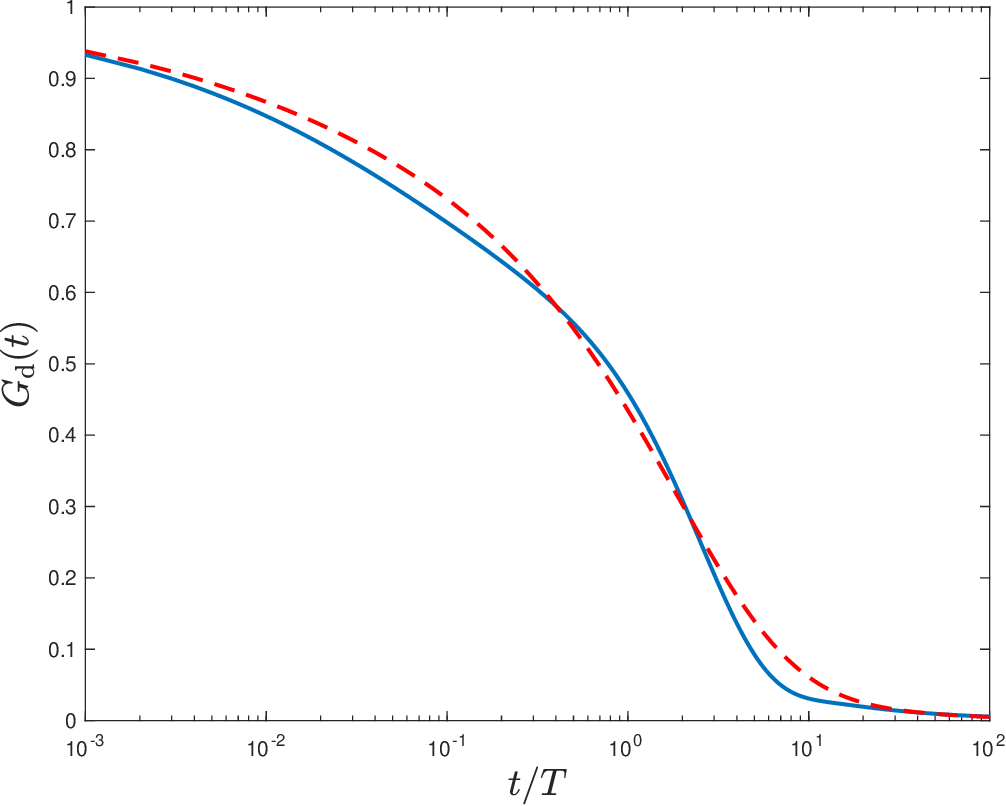}
    \caption{Comparison of the relaxation modulus for the variable-order fractional Maxwell model with: {\em (a)} an exponential-type transition function [continuous line]; {\em (b)} a Mittag-Leffler-type transition function [dashed line]. Assuming: $\alpha_1 = 0.4$, $\alpha_2=0.8$, $c=1$, $a_1 = 1$, and the Mittag-Leffler parameter (see Eq.~\eqref{eq:alpha-ML-trans}) $\beta = 0.5$.}
    \label{fig:Maxwell-ML-Exp-comparison}
\end{figure}

In Fig.~\ref{fig:Gd-Maxwell} and \ref{fig:Jd-Maxwell} we have plotted the numerical evaluation of the time-representation of the dimensionless relaxation modulus and creep compliance, respectively, for the variable-order fractional Maxwell model as a function of the rescaled time $t/T$. These figures clearly show that the model starts-off as a fractional Maxwell model of order $\alpha_1$ at $t=0$ and smoothly approaches a new ``phase'' given by a fractional Maxwell model of order $\alpha_2$ at late time ({\em i.e.}, as $t\to+\infty$). This transition is controlled by an exponential-type order function $\alpha(t)$ as in Eq.~\eqref{eq:alpha-exp-trans}. Replacing the exponential-type transition function with a Mittag-Leffler-type one does not lead to major qualitative differences in the overall behaviour of the dimensionless material functions, as shown in Fig.~\ref{fig:Maxwell-ML-Exp-comparison}.

\begin{figure}[t]
    \centering
    \includegraphics[scale=0.8]{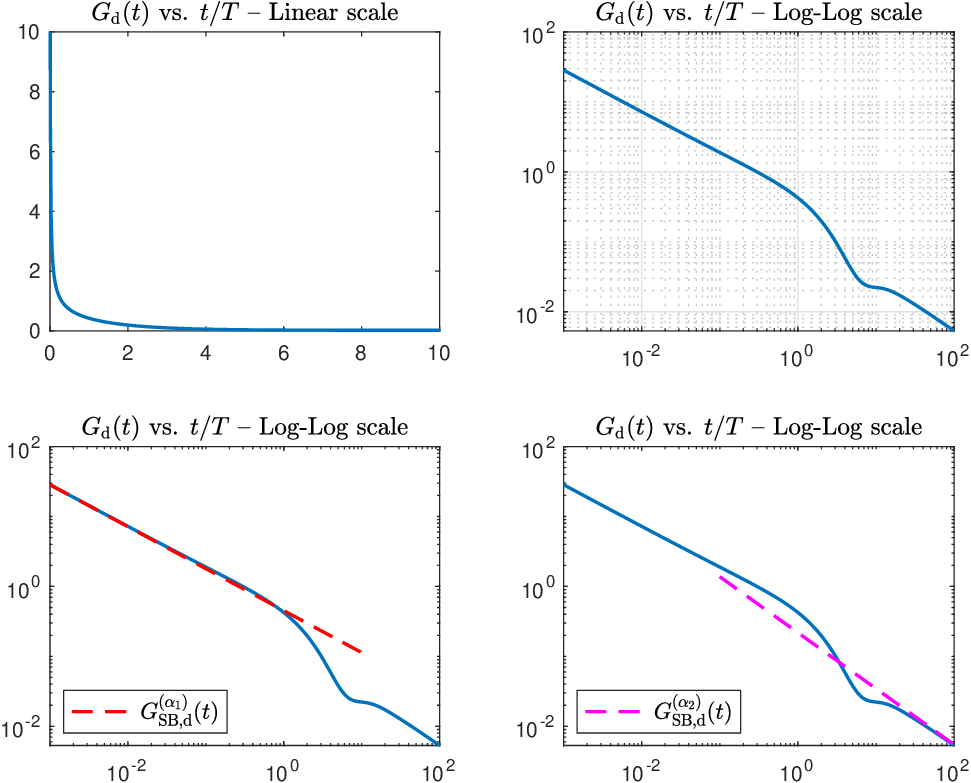}
    \caption{Dimensionless relaxation modulus ${G}_{\rm d}(t)$ [continuous line] as a function of $t/T$ for the variable-order Scott Blair model given an exponential-type transition $\alpha(t)$ (see Eq.~\eqref{eq:alpha-exp-trans}) with $\alpha_1 = 0.6$, $\alpha_2=0.8$, $c=2$, and $a_1 = 1$.}
    \label{fig:Gd-SB}
\end{figure}
\begin{figure}[t]
    \centering
    \includegraphics[scale=0.8]{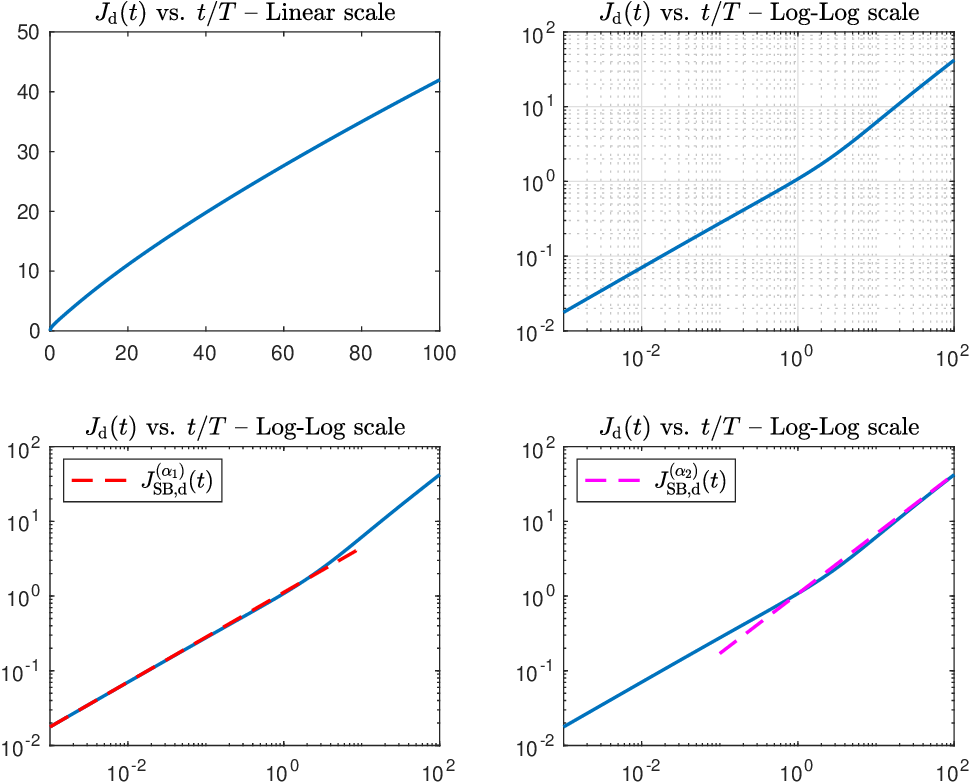}
    \caption{Dimensionless creep compliance ${J}_{\rm d}(t)$ [continuous line] as a function of $t/T$ for the variable-order Scott Blair model given an exponential-type transition $\alpha(t)$ (see Eq.~\eqref{eq:alpha-exp-trans}) with $\alpha_1 = 0.6$, $\alpha_2=0.8$, $c=2$, and $a_1 = 1$.}
    \label{fig:Jd-SB}
\end{figure}
Fig.~\ref{fig:Gd-SB} and \ref{fig:Jd-SB} address the case of the variable-order Scott Blair model, corresponding to $a_1 = 0$ in Eq.~\eqref{eq:scarpi-maxwell}, and again these results confirm the discussion already presented for the case of the more general variable-order fractional Maxwell model.
\newpage
\section{Comments on previous literature}
\label{sec:massacro}
In this section we briefly discuss a recent attempt \cite{cinesata} to apply Scarpi's fractional calculus in viscoelasticity and viscoplasticity that, while being of potential interest in the general sense, suffers of some conceptual shortcomings.

First, and most importantly, the choice of a linear transition function $\alpha(t) = a \, t + b$ (assuming a dimensionless time $t$ and dimensionless constants $a, b>0$) is not consistent with the formalism underlying Scarpi's operators. Indeed, the transition function is supposed to be $0 < \alpha (t) < 1$ at all times and satisfy {\em (i--iii)} in Sec.~\ref{sec:2}. However, $\alpha(t) = a \, t + b$ is unbounded on $[0,+\infty)$ and it violates the constraint $\alpha (t) < 1$ for all $t\geq(1-b)/a$. A proper reformulation of the problem would require a transition function $\alpha (t)$ such that
$$
\alpha (t) = \left\{ 
\begin{aligned}
    0 \, ,& \qquad \mbox{if} \,\, t<0 \, ,& \\
    a \, t + b  \, ,& \qquad \mbox{if} \,\, 0\leq t \leq \frac{1-b}{a} - \delta \, ,& \\
    1 - \delta \, a \, ,& \qquad \mbox{if} \,\, t>\frac{1-b}{a} - \delta \, ,&
\end{aligned}
\right.
$$
with $a>0$, $0<b<1$, and $\delta>0$ a small real constant. However, this drastically affects the corresponding Laplace transform, that now yields
$$
A(s) = \frac{b}{s} + \frac{a}{s^2} \left\{ 1 - 
\exp \left[- \left( \frac{1-b}{a} - \delta \right) \, s \right]
\right\} \, ,
$$
which is significantly different from the $A(s) = \frac{b}{s} + \frac{a}{s^2}$ considered in \cite{cinesata}.

Second, the fractionalisation procedure for the Scott Blair model presented in \cite{cinesata}, although interesting in itself, makes the model nonlinear. However, it is worth noting that basically the same idea was put forward by Gianbattista Scarpi in his 1972 fundamental paper \cite{Scarpi1972a}. The objective of our work is instead to provide a general prescription for implementing the Scarpi fractional calculus in {\em linear} viscoelasticity.

Lastly, although something similar to the fractionalisation proposed in this work seem to have been discussed in \cite{cinesata} concerning the fractional Zener model (albeit still considering a linear transition function in time that suffers of the same shortcomings discussed above), the dimensions of the model do not seem to be correct, differently from the nonlinear Scott Blair and the nonlinear visco-plastic models.

The scope of this comment is not to take away any merit to the investigation in \cite{cinesata}, that might have its own realm of applicability that is however beyond the scope of our study. This is just meant to clarify some technical misconceptions concerning the Scarpi fractional calculus and its regime applicability.
\section{Conclusions}
\label{sec:conc}
In the recent literature, some attention has been devoted to the study of applications of variable-order fractional derivatives in viscoelasticity (see {\em e.g.}, \cite{liu}, \cite{meng}, and references therein). However, there exists a plethora of different and nonequivalent approaches to 
variable-order fractional calculus and in the vast majority of applications of such operators the choice of the time-dependent order function $\alpha(t)$ is motivated in terms of ``best-fitting procedures'' rather than physical intuition.

In this work we have built upon the previous literature \cite{cinesata} and provided for the first time a consistent formulation of a {\em variable-order fractional linear theory of viscoelasticity} based on Scarpi's approach to fractional calculus. Moreover, we have suggested a theoretical formulation of the problem based on the idea that the evolution of the order of the fractional derivative can be governed by a specific differential equation, related to the physical process.

In Sec.~\ref{sec:2} we have reviewed the essentials about the mathematical formulation of Scarpi's operators, with particular regard for Exponential-type and Mittag-Leffler-type transition functions $\alpha (t)$.

In Sec.~\ref{sec:3} we have introduced the fractionalisation procedure aimed at consistently reformulating fractional linear viscoelasticity in terms of variable-order derivatives. We considered as a toy example the case of the {\em fractional Maxwell model}, that incidentally includes also the Scott Blair model as a particular case, and computed the associated creep compliance and relaxation modulus. Sec.~\ref{sec:3-1} provides a physical motivation for the use of variable-order fractional calculus for physical applications and discusses the interpretation of variable-order scenarios as a system of coupled evolution equations for both the order and the physical observables. These evolution equations, capturing the full complexity of the (macroscopic) system, should therefore concurrently emerge from a theoretical (microscale) constructed out of fundamental principles. This is perhaps the most important and novel idea that we wish to highlight with this study. For this reason, we focused our attention on transition functions governed by relaxation processes that appear to be almost ubiquitous in all applied sciences. In Sec.~\ref{sec:3-1} we then evaluated numerically these material functions taking advantage of the numerical methods discussed in \cite{GarrappaGiustiMainardi2021,GarrappaGiusti2023}) for Exponential-type and Mittag-Leffler-type transition functions $\alpha (t)$.

In Sec.~\ref{sec:massacro} we provide a critical assessment of some shortcoming of a previous attempt to implementing the Scarpi fractional calculus in viscoelasticity and viscoplasticity \cite{cinesata}. In particular, we discuss how the mathematical formulation presented here preserves the linearity of the model post-fractionalisation and the need for a careful manipulation of the transition functions so that the problem remains well-defined within Scarpi's approach. Additionally, we present our analysis for the fractional Maxwell model in order to remain fully complementary to \cite{cinesata}.

To conclude, Scarpi's approach has opened a window on brand new applications of variable-order fractional calculus in mechanical systems that deserve to be extensively explored in the future. Perhaps, \cite{Zorica} and references therein could serve as inspiration for future investigations.

\section*{Acknowledgments}
We are grateful to Francesco Mainardi for useful discussions. The work of R.~Garrappa is partially supported by the MUR under the PRIN-PNRR2022 project n. P2022M7JZW and by the INdAM under the GNCS Project E53C22001930001. The work of A.~Mentrelli is partially supported by the MUR under the PRIN2022 PNRR project n.~P2022P5R22A. The work of I.~Colombaro is partially supported by the INdAM-GNFM Young Researchers Project 2023, CUP\_E53C22001930001. This work has been carried out in the framework of the activities of the Italian National Group of Mathematical Physics [Gruppo Nazionale per la Fisica Matematica (GNFM), Istituto Nazionale di Alta Matematica (INdAM)].
%
%

%
%
%
\end{document}